\documentclass[manuscript,preprintnumbers,amsmath,amssymb]{revtex4}
\usepackage{graphicx}
\begin{document}

\title{A confirmation of agreement of different approaches for scalar gauge-invariant metric perturbations during inflation}
\author{$^{1,2}$ Mariano Anabitarte\footnote{E-mail address: anabitar@mdp.edu.ar} and $^{1,2}$ Mauricio
Bellini \footnote{E-mail address: mbellini@mdp.edu.ar}}

\address{$^{1}$ Departamento de F\'{\i}sica, Facultad de Ciencias Exactas y
Naturales, Universidad Nacional de Mar del Plata, Funes 3350,
(7600) Mar del Plata,
Argentina.\\
$^2$ Consejo Nacional de Investigaciones Cient\'{\i}ficas y
T\'ecnicas (CONICET). }

\begin{abstract}
We revisit an extension of the well-known formalism for
gauge-invariant scalar metric fluctuations, to study the spectrums
for both, the inflaton and gauge invariant (scalar) metric
fluctuations in the framework of a single field inflationary model
where the quasi-exponential expansion is driven by an inflation
which is minimally coupled to gravity. The proposal here examined
is valid also for fluctuations with large amplitude, but for
cosmological scales, where vector and tensor perturbations can be
neglected and the fluid is irrotacional.
\end{abstract}


\maketitle
\section{Introduction}

The big bang theory has provided a remarkably successful
description of the evolution of the universe and now rests upon
solid observational foundations. Over the last decades
cosmologists have also drawn up architectural sketches of a model
which may account for the origin and evolution of structure within
this framework: primordial perturbations from an early epoch of
inflation, which subsequently grow by gravitational instability to
form galaxies and larger structures. The standard picture of
inflation introduced in 1981\cite{Guth} relied on a scalar field,
called the inflaton, which during inflation was assumed to have no
interaction with any other field. The inflationary scenario
postulates that the universe underwent a phase of very rapid,
accelerated expansion in its distant past. Observations have
provided strong support for the paradigm. However, despite this
success, the mechanism which drove the inflationary expansion has
yet to be indentified. A multitude of inflationary models
involving a broad range of energy scales have been discussed in
the literature, including chaotic inflation\cite{chaotic}, warm
inflation\cite{wi}, stochastic inflation\cite{si}, fresh
inflation\cite{fi}, brane inflation\cite{bi}, STM
inflation\cite{STM}, and many others.

The scalar fluctuations of the metric are associated with density
perturbations. These are the spin-zero projections of metric
perturbations and were induced by the vacuum fluctuations of the
inflaton field during inflation. Furthermore, they played a
crucial role in the generation of primordial inhomogeneities which
gave rise to the large scale structure of the present day universe
as well as the observed anisotropies of the cosmic microwave
background of radiation. Gauge invariance guarantees that equation
for the fluctuations of the geometry do not change when moving
from one coordinate system to other. This allows us to formulate
the problem of the evolution for the amplitude of scalar metric
perturbations around the Friedmann-Robertson-Walker (FRW) universe
in a coordinate-independent manner at every moment in time. The
issue of gauge invariance becomes critical when attempt to analyze
how the scalar metric perturbations produced in the early universe
influence a globally flat, isotropic and homogeneous universe on
super Hubble scales. Space-time fluctuations can also lead to
decoherence of matter waves\cite{Power}. In the infrared (IR)
sector these fluctuations can be represented by a coarse-grained
field, which describes a stochastic dynamics\cite{Bellini}. With
respect to perturbative approaches, second order gauge invariant
perturbations quantities have been calculated in \cite{second} and
third order perturbations are supposed to be
negligible\cite{third} during inflation, but not for large
fluctuations.

In this work we examine a perturbed FRW metric using the proposal
developed in\cite{kolb}, which is an extension of the well known
linearized line element $ds^2=(1+2\psi)
dt^2-a(t)^2\,(1-2\psi)\,d\vec{r}^2$, for a longitudinal gauge. Our
results show a coincidence of different approaches for small
fluctuations; the formalism developed in \cite{kolb} and the
standard method.

\section{Formalism}

We consider a scalar field $\varphi$ which is minimally coupled to
gravity. The action of the system is
\begin{equation}
I= {\Large{\int}} d^4x\,\sqrt{-g} \left[ \frac{\cal{R}}{16\pi G}
+\frac{1}{2} g^{\mu\nu} \varphi_{,\mu}\varphi_{,\nu} -
V(\varphi)\right],
\end{equation}
where $g= -e^{-4\psi} a^6(t)$ is the determinant of the covariant
metric tensor with components $g_{\mu\nu}$ ($\mu,\nu$ run from $0$
to $3$) and $V(\varphi)$ is the potencial related to the inflaton
field. To study the gauge invariant scalar metric fluctuations
$\psi\left(x^{\alpha}\right)$, we propose the following perturbed
metric
\begin{equation}\label{me}
dS^2= e^{2\psi}\,dt^2 - a^2(t) e^{-2\psi} \,d\vec{r}^2,
\end{equation}
where $d\vec{r}^2 = dx^2+dy^2+dz^2$ and $\psi(t,\vec{r})$ is the
scalar metric fluctuation. The metric (\ref{me}) is the perturbed
version of the background FRW one, which is spatially flat,
isotropic, homogeneous and has a scalar curvature $\bar{\cal{
R}}=6\left[{\ddot a\over a} +\left({\dot a\over
a}\right)^2\right]$. This metric describe non-perturbative
gravitational fluctuations on cosmological scales, on which vector
and tensor perturbations of the metric can be neglected and the
fluid can be considered as irrotacional. Furthermore, using the
continuity equation on large scales
\begin{equation}
\frac{\partial\rho}{\partial\tau} = -3{\cal{H}} \left( \rho + P
\right),
\end{equation}
where ${\cal{H}} = {d\over d\tau}\left[{\rm{ln}}\left( a(t)\,
e^{-\psi}\right)\right]$ and $d\tau =e^{\psi} dt$. One can show
that exists a conserved quantity in time at any order in
perturbation theory
\begin{equation}
f = {\rm{ln}} \left( a e^{-\psi}\right) + \frac{1}{3}
{\Large{\int}}^{\rho} \frac{d\rho'}{\left(P'+\rho'\right)}.
\end{equation}
Considering the invariant $\omega$, which characterizes the
equation of state $P = \omega\,\rho$. The perturbation $\delta
f=-\psi +{1\over 3(1+\omega)}
{\rm{ln}}\left(\rho/\bar\rho\right)$, is a gauge-invariant
quantity representing the non-linear extension of the curvature
perturbation for adiabatic fluids on uniform energy density
hypersurfaces on superhorizon scales\cite{kolb}. Here, ($\bar P,\,
\bar\rho$) denote the background pressure and energy density and
($P,\,\rho$) denote pressure and energy density on the perturbed
metric (\ref{me}). The scalar curvature of this metric is
\begin{equation}
{\cal{R}} = 6 e^{-2\psi} \left[ \left[\frac{\ddot a}{a} +
\left(\frac{\dot a}{a}\right)^2\right] - \ddot\psi e^{4\psi} -
5\dot\psi \left(\frac{\dot a}{a}\right) + \frac{e^{4\psi}}{3a^2}
\left[\nabla^2\psi - \left(\vec{\nabla}\psi\right)^2\right]
+3\dot\psi^2\right].
\end{equation}
Now we shall use the Lagrangian formalism to describe the dynamics
of $\varphi$ and $\psi$. The equation of motion for $\varphi$ is
\begin{equation}\label{ec1}
\ddot\varphi + \left[3\left(\frac{\dot a}{a}\right)-4\dot\psi
\right] \dot\varphi - \frac{1}{a^2} e^{4\psi} \nabla^2\varphi +
e^{2\psi} V'(\varphi)=0,
\end{equation}
where $V'(\varphi)\equiv {dV\over d\varphi}$. The equation
(\ref{ec1}) is an operatorial one because $\varphi$ is considered
as a quantum scalar field. The equation of motion for $\psi$ is
\begin{equation}\label{ec2}
\left[\frac{\partial {\cal{ R}}}{\partial\psi} -
\frac{\partial}{\partial x^{\mu}} \frac{\partial
\cal{R}}{\partial\psi_{,\mu}}\right] - 2 {\cal{R}}
-\frac{1}{\sqrt{-g}} \frac{\partial \sqrt{-g}}{\partial t}
\frac{\partial {\cal R}}{\partial \dot\psi} = 32\pi G \left[
e^{-2\psi} \dot\varphi^2 -
e^{2\psi}\frac{1}{a^2}\left(\vec{\nabla}\varphi\right)^2-V(\varphi)\right],
\end{equation}
such that both, $\varphi$ and $\psi$ comply with the commutation
relations
\begin{eqnarray}
&& \left[\varphi(t,\vec r),\Pi_{\varphi}(t,\vec{r´})\right] = i\,
\delta^{(3)}\left(\vec{r}-\vec{r´}\right), \\
&& \left[\psi(t,\vec r),\Pi_{\psi}(t,\vec{r´})\right] = i\,
\delta^{(3)}\left(\vec{r}-\vec{r´}\right),
\end{eqnarray}
$\Pi_{\varphi}$ and $\Pi_{\psi}$ being respectively the conjugate
momentums for $\varphi$ and $\psi$:
\begin{displaymath}
\Pi_{\varphi} = \frac{\partial L}{\partial \dot\varphi}, \qquad
\Pi_{\psi} = \frac{\partial L}{\partial \dot\psi},
\end{displaymath}
such that $L = \sqrt{-g}\,\left[{{\cal R}\over 16\pi G} + {1\over
2} g^{\mu\nu} \varphi_{,\mu}\varphi_{,\nu} - V(\varphi)\right]$ is
the density Lagrangian of the system. To complete the description
of the dynamics, we need to write the Einstein equations
$G_{\alpha\beta} =-8\pi G T_{\alpha\beta}$. Taking into account
cartesian coordinates, the diagonal Einstein equations are
\begin{eqnarray}
&& -\frac{2}{a^2} e^{4\psi} \nabla^2\psi + 6 H \dot\psi
-3\dot\psi^2 + \frac{e^{4\psi}}{a^2}
\left(\vec{\nabla}\psi\right)^2 + 3H^2 \nonumber \\
&& = -8\pi G
\left[\frac{\dot\varphi^2}{2}+e^{4\psi}\frac{1}{2a^2}\left(\vec\nabla\varphi\right)^2+V(\varphi)e^{2\psi}\right],\label{ei1}\\
&& -8 \pi G \left[\frac{3 e^{-4\psi}}{2} \dot\varphi^2 -\frac{1}{2
a^2} \left(\vec\nabla\varphi\right)^2- 3 e^{-2\psi}V\right]
=-\frac{2}{a^2} \left(\vec{\nabla}\psi\right)^2- 24 H\dot\psi
e^{-4\psi} \nonumber
\\
&&  +15\dot\psi^2 e^{-4\psi} + 6 e^{-4\psi}\left(\frac{\ddot
a}{a}\right) - 6\ddot\psi e^{-4\psi} -3H^2 e^{-4\psi} \label{ei2},
\end{eqnarray}
where $G^{\alpha}_{\,\,\beta}=R^{\alpha}_{\,\,\beta}-{1\over 2}
{\cal{R}} g^{\alpha}_{\,\,\beta}$ is the Einstein tensor and
$T^{\alpha}_{\,\,\beta}=\varphi^{,\alpha}\varphi_{,\beta}-g^{\alpha}_{\,\,\beta}
\left[{1\over 2} \varphi^{,\rho}\varphi_{,\rho} -
V(\varphi)\right]$ is the Energy - Momentum tensor for a scalar
field. On the other hand, the non-diagonal Einstein equations have
the form
\begin{equation}\label{ec4}
\frac{1}{a}\frac{\partial}{\partial x^i}\left[
\frac{\partial}{\partial t} \left(a\psi\right)\right] -
\frac{\partial\psi}{\partial t} \frac{\partial\psi}{\partial x^i}
= 4\pi \,G\,\frac{\partial\varphi}{\partial t}
\frac{\partial\varphi}{\partial x^i},
\end{equation}
so that, using the eq. (\ref{ec4}) in eqs. (\ref{ei1}) and
(\ref{ei2}), we obtain the exact equation of motion for $\psi$
\begin{equation}\label{mot2}
\ddot\psi + 7 H \dot\psi - \frac{e^{4\psi}}{a^2} \nabla^2\psi  -
4\dot\psi^2 -
\frac{5e^{4\psi}}{3a^2}\left(\vec{\nabla}\psi\right)^2 + 8\pi G V
(\varphi) e^{2\psi} = -\frac{8\pi G}{3a^2}
\left(\vec{\nabla}\varphi\right)^2 ,
\end{equation}
which describes the dynamics of $\psi$ with arbitrary amplitude.
However, it is very difficult to solve these equations in an exact
manner. Notice that we have used the Einstein equations, and not
the Lagrange one, to obtain the dynamics of $\psi$. One could make
the inverse procedure, because both manners to work are
equivalent. However, in this case the calculations with the
Einstein equations are more simple.\\

\section{Linear approximation}

In the weak field limit, it is sufficient to make a linear
approximation on the scalar metric perturbations, so that one can
write $e^{\pm 2\psi(x^{\alpha})} \simeq 1 \pm 2\psi(x^{\alpha})$
in the exact equations of motion (\ref{ec1}) and (\ref{mot2}), for
$\psi$ and $\varphi$. In this limit the metric (\ref{me})
preserves gauge invariance and the linearized line element
\begin{displaymath}
dS^2=(1+2\psi )\, dt^2 - a(t)^2 \,(1-2\psi ) d\vec{r}^2,
\end{displaymath}
takes the form of a longitudinal gauge so that coordinate
transformations induce difeomorfism transformations\cite{bardeen}.

The equation of motion for the inflaton in its exact form is given
by (\ref{ec1}). However, in the weak field limit we can make the
semi-classical approximation $\varphi(x^{\alpha}) = \left<E
|\varphi|E\right> +\phi(x^{\alpha})$. Here, $\left<E
|\varphi|E\right>=\phi_c(t)$ is the expectation value of $\varphi$
evaluated on the quantum state $\left.|E\right>$. Furthermore, in
this limit the quantum fluctuations of the inflaton field are
considered to be very small and $\left<E|\phi|E\right>
=\left<E|\dot\phi|E\right>=0$. With this representation we obtain
the following dynamical equations for the fluctuations of the
inflaton field $\phi$ and the classical field $\phi_c$:
\begin{eqnarray}
&&\ddot\phi +3 \left(\frac{\dot a}{a}\right) \dot\phi -
\frac{1}{a^2} \nabla^2\phi + V''(\phi_c) \,\phi = 4 \dot \psi
\dot\phi_c + 2
\psi V'(\phi_c), \\
&& \ddot\phi_c + 3 \left(\frac{\dot a}{a}\right) \dot\phi_c +
V'(\phi_c) =0.
\end{eqnarray}
The background Friedmann equations are
\begin{equation}
3H^2_c = 8\pi G \left[\frac{\dot\phi^2_c}{2} + V(\phi_c)\right],
\end{equation}
where $V(\phi_c)\equiv \left. V(\varphi)\right|_{\phi_c}$ is the
scalar potential evaluated on the classical background field
$\phi_c(t)$ and the Hubble parameter with back-reaction effects
included is
\begin{equation}\label{ag}
H = \frac{\dot a}{a} \simeq H_c(t) + \frac{4\pi G}{3 H_c} \left<
E\right| \frac{\dot\phi^2}{2} + \frac{(\vec\nabla\phi)^2}{2 a^2} +
\sum_{n=1}^{\infty} \frac{V^{(n)}(\phi_c)}{n!} \,
\phi^{n}(x^{\alpha})\left|E\right>.
\end{equation}
When the metric fluctuations are small it is sufficient to make
$H_c\simeq {\dot a\over a}$, because the last term in the right
hand of the expression (\ref{ag}) is negligible with respect to
the first one. This approximation is valid only on large scales,
which are super Hubble scales during the inflationary epoch.
Furthermore, the primer denotes the derivative with respect to
$\varphi$, such that $ V'(\phi_c) \equiv \left. {d V(\varphi)\over
d\varphi} \right|_{\phi_c}$.

Furthermore, the Einstein equations (\ref{ec4}) now hold
\begin{equation}\label{18}
\frac{1}{a} \frac{\partial^2}{\partial x^i \partial t}\left[a \,
\psi\right] = 4\pi G \, \frac{\partial}{\partial
x^i}\left[\dot\phi_c\,\phi \right],
\end{equation}
from which (once we have taken into account that
$\left<E|\phi|E\right>=0)$ we obtain that the evolution for the
expectation value of $ \psi$ goes as
\begin{equation}
\left< E|\psi|E\right > \sim a^{-1},
\end{equation}
which decreases with the inverse of the scale factor of the
universe. Finally, the linearized dynamics of $\psi$ can be
obtained from the Einstein equations (\ref{ei1}) and (\ref{ei2}):
$\delta G^{\mu}_{\,\,\nu} = -8\pi G \,\delta T^{\mu}_{\,\,\nu}$
\begin{equation}\label{equ}
\ddot\psi + \left[H-2\frac{\ddot\phi_c}{\dot\phi_c}\right]
\dot\psi - \frac{1}{a^2}\nabla^2\psi + 2\left[\dot H -
\frac{\ddot\phi_c}{\dot\phi_c} H \right]\psi =0,
\end{equation}
which, as one expects, is the same as the equation obtained
in\cite{Mukhanov}. Note that the equation (\ref{equ}) is the
equation (\ref{mot2}) once linearized, with the constriction
(\ref{18}).

\section{An example}

In this section we shall illustrate the formalism in the
linearized approximation, when the expansion is governed by a
power-law expansion $a =\beta\, t^{p}$. In this case the Hubble
parameter is given by $H=p/t$ and the classical field $\phi_c(t)$
is
\begin{equation}
\phi_c(t) = \phi_0 \left[1-{\rm ln}\left(\frac{H_0
t}{p}\right)\right],
\end{equation}
where the power $p$ is
\begin{displaymath}
p = 4\pi \, G \,\phi^2_0,
\end{displaymath}
$\phi_0$ being the value of $\phi_c(t_0)$ when inflation begins
(i.e., at $t=t_0$). The equation of state is given by $\omega =
{\bar P\over \bar\rho}={-\left(p-2/3\right)\over p}$, $\omega$
being an invariant. Furthermore, the classical potential
$V(\phi_c)$ is
\begin{equation}
V(\phi_c) = \frac{3 H^2_0}{8\pi G}\left(\frac{3p-1}{3p}\right)\,
e^{2\phi_c/\phi_0}.
\end{equation}
The solution for the $\psi$-modes, once normalized, are
\begin{equation}
\xi_k(t) = \frac{\sqrt{\pi}}{2} \sqrt{\frac{t}{(p-1)}} \, {\cal
H}^{(2)}_{\mu}\left[\frac{k \,t^{1-p}}{(p-1) \beta}\right] \times
\left(\frac{t}{t_0}\right)^{-(p+2)/2},
\end{equation}
where ${\cal H}^{(2)}_{\mu}\left[x(t)\right]$ is the second kind
Hankel function with $\mu={p+1\over 2(p-1)}$. Using the small
argument Hankel functions limit, we obtain that these modes have
the following expression on super Hubble scales:
\begin{equation}
\left.\xi_k(t)\right|_{k \gg 1/(a H)} \simeq
i\,\,\sqrt{\frac{\pi}{4(p-1)}}\,\,
\Gamma\left[\frac{(p+1)}{2(p-1)}\right]
\left[\frac{2(p-1)\beta}{\pi}\right]^{\frac{(p+1)}{2(p-1)}}
k^{\frac{-(p+1)}{2(p-1)}},
\end{equation}
which is independent of time. The equation of motion for the modes
of the inflaton field on cosmological scales can be approximated
to
\begin{eqnarray}
&&\ddot{\tilde\xi}_k(t)+ \frac{3p}{t}\dot{\tilde\xi}_k(t) +
\left[\frac{k^2}{\beta^2\, t^{2p}} +\frac{4\left(6\pi G \phi^2_0
-1\right)}{t^2}\right] \tilde\xi_k(t) \nonumber \\
&& = {\rm i} \left[\frac{\left(3p-2\right) \phi_0}{t^2}\right]
\sqrt{\frac{\pi}{p-1}} \Gamma\left(\frac{p+1}{2(p-1)}\right)
\left[ \frac{2\beta(p-1)}{\pi}\right]^{\frac{(p+1)}{2(p-1)}} \,
k^{-\frac{(p+1)}{2(p-1)}},
\end{eqnarray}
which, on super Hubble scales, has the solution
\begin{eqnarray}
\tilde\xi_k(t) & \simeq & t^{\frac{1}{2}(1-3p)} \nonumber
\\
& \times & \left\{ A\,\, {\cal J}_{\nu_2}\left[x(k,t)\right] + B
\,\, {\cal Y}_{\nu_2}\left[x(k,t)\right]  - {\rm i} \gamma \beta
\sqrt{\frac{\pi}{p-1}} \Gamma\left(\frac{p+1}{2(p-1)}\right)
\left[\frac{2\beta(p-1)}{\pi}\right]^{\frac{p+1}{2(p-1)}}
\,\,k^{\frac{1-3p}{2(p-1)}} \right.\nonumber \\
&\times & \left[ {\cal J}_{\nu_2}\left[x(k,t)\right] \Large{\int}
\frac{dt\,\, t^{\frac{5}{2}(p-1)} \,{\cal
Y}_{\nu_2}\left[x(k,t)\right] }{\left[ {\cal
J}_{\nu_2}\left[x(k,t)\right] \, {\cal
Y}_{\nu_1}\left[x(k,t)\right] - {\cal
Y}_{\nu_2}\left[x(k,t)\right]\,{\cal
J}_{\nu_1}\left[x(k,t)\right]\right]}\right.\nonumber \\
&-& \left.\left.\left[ {\cal
Y}_{\nu_2}\left[x(k,t)\right]\,\Large{\int} \frac{dt\,\,
t^{\frac{5}{2}(p-1)} \,{\cal J}_{\nu_2}\left[x(k,t)\right]
}{\left[ {\cal J}_{\nu_2}\left[x(k,t)\right] \, {\cal
Y}_{\nu_1}\left[x(k,t)\right]{\cal
Y}_{\nu_2}\left[x(k,t)\right]\,{\cal
J}_{\nu_1}\left[x(k,t)\right]\right]}\right]\right]\right\},
\end{eqnarray}
where $\gamma = 3(p-2) \phi_0$, $3p > m^2 = 4(6\pi G \phi^2_0
-1)$, $\nu_2={\sqrt{9p^2-6 p+1-4m^2}\over 2(p-1)}$, $\nu_1 =
1-\nu_2$ and $x(k,t)= {k t^{(1-p)}\over \beta(p-1)}$. For $x(t)
\ll 1$ this solution can be written as
\begin{eqnarray}
\tilde\xi_k(t) &\simeq & - \frac{B}{\pi} \Gamma\left(\nu_2\right)
\left(\frac{k}{2\beta(p-1)}\right)^{-\nu_2} \, t^{-(1-p)\nu_2
+\frac{1}{2}(1-3p)} \nonumber \\
&+ & {\rm i} \beta (3p-2) \phi_0
\Gamma\left(\frac{p+1}{2(p-1)}\right)
\,\left[\frac{2\beta(p-1)}{\pi}\right]^{\frac{p+1}{2(p-1)}}
\nonumber \\
& \times & \left[\frac{\Gamma\left(\nu_2+1\right)
\,t^{-(\nu_1+1)(1-p)}}{2\Gamma\left(\nu_1+1\right)
\left[(\nu_2-\nu_1)(p-1)\right]}
\left(\frac{k}{2\beta(p-1)}\right)^{\nu_2-\nu_1}\right. \nonumber
\\
&\times& \left.\Large\sum_{n=0}^{\infty}
\,\frac{1}{\left[\frac{5p-3+2\nu_1(p-1)}{4(\nu_2-\nu_1)(1-p)}+1\right]}\left[-\frac{\Gamma(\nu_2+1)
\Gamma(\nu_1)}{\Gamma(\nu_2)
\Gamma(\nu_1+1)}\,x(k,t)\right]^{2n(\nu_2-\nu_1)}\right.\nonumber \\
& + &
\left.\frac{\Gamma(\nu_2)\,t^{-(2\nu_2+1)(1-p)}}{2\Gamma(\nu_2+1)\left(\nu_2-\nu_1\right)\left(p-1\right)}
\left(\frac{k}{2\beta(p-1)}\right)^{-2\nu_2}\right.\nonumber \\
&\times & \left.\Large\sum_{n=0}^{\infty}
\,\frac{1}{\left[\frac{5p-3+2\nu_2(p-1)}{4(\nu_2-\nu_1)(p-1)}+1\right]}\left[-\frac{\Gamma(\nu_1+1)
\Gamma(\nu_2)}{\Gamma(\nu_1)
\Gamma(\nu_2+1)}\,x(k,t)\right]^{2n(\nu_1-\nu_2)}\right],
\end{eqnarray}
where $-1/2 < \nu_1 <1/2$ and $1/2<\nu_2 <3/2$.

Now we are interested in obtaining the spectrum of the $\psi$ and
$\varphi$ squared-fluctuations. Their spectrums ${\cal
P}_{\varphi}(k,t)$ and ${\cal P}_{\psi}(k,t)$ on cosmological
scales, are given respectively by the expressions
\begin{eqnarray}
\left.{\cal P}_{\varphi}(k,t)\right|_{IR} &=& \frac{k^3}{2\pi^2}
\left(\tilde\xi_k \tilde\xi^*_k \right)\simeq \frac{k^3}{2\pi^2}
\left[ A^2_1(t) k^{-2\nu_2} + \left[ B_1(t) k^{2(\nu_2-\nu_1)}
\Large\sum_{n=0}^{\infty} \frac{\left[\alpha_1
\,\,x(k,t)\right]^{2n(\nu_2-\nu_1)}}{\left(v_1+1\right)}
\right.\right.\nonumber \\
& + & \left.\left. B_2(t) k^{2(\nu_1-\nu_2)}
\Large\sum_{n=0}^{\infty} \frac{\left[\alpha_2
\,\,x(k,t)\right]^{2n(\nu_1-\nu_2)}}{\left(v_2+1\right)}\right]^2\right], \label{p1}\\
\left.{\cal P}_{\psi}(k,t)\right|_{IR} & = & \frac{k^3}{2\pi^2}
\left(\xi_k \xi^*_k\right) \simeq \frac{1}{4\pi(p-1)}
\,\,\Gamma\left(\frac{p+1}{2(p-1)}\right)^2 \,
\left[\frac{2(p-1)\beta}{\pi}\right]^{\frac{p+1}{p-1}} \,\,
k^{3-\frac{p+1}{p-1}}, \label{p2}
\end{eqnarray}
where
\begin{eqnarray}
&& B_1(t)= \frac{\beta (3p-2)\phi_0 \Gamma(\nu_2+1)
\Gamma\left(\frac{p+1}{2(p-1)}\right)\left[2\beta(p-1)\right]^{\frac{p+1}{2(p-1)}+\nu_2-\nu_1}}{2\pi^{\frac{p+1}{2(p-1)}}
\Gamma(\nu_1+1)\left[(\nu_2-\nu_1)(p-1)\right]}
t^{-(\nu_1+1)(1-p)}, \\
&& B_2(t)= \frac{\beta (3p-2)\phi_0 \Gamma(\nu_2)
\Gamma\left(\frac{p+1}{2(p-1)}\right)\left[2\beta(p-1)\right]^{\frac{p+1}{2(p-1)}+2\nu_2}}{2\pi^{\frac{p+1}{2(p-1)}}
\Gamma(\nu_2+1)\left[(\nu_2-\nu_1)(p-1)\right]}
t^{-(2\nu_2+1)(1-p)}, \\
&& A= -\frac{B\Gamma(\nu_2)\,t^{\nu_2(p-1)+(1-3p)/2}}{\pi \left[2\beta(p-1)\right]},\\
&& \alpha_1 =
-\frac{\Gamma(\nu_2+1)\,\Gamma(\nu_1)}{\Gamma(\nu_2)\,\Gamma(\nu_1+1)},
\\
&& \alpha_2 =
-\frac{\Gamma(\nu_1+1)\,\Gamma(\nu_2)}{\Gamma(\nu_2)\,\Gamma(\nu_2+1)},
\\
&& v_1= \frac{5p-3+2\nu_1(p-1)}{4(\nu_2-\nu_1)(1-p)}, \\
&& v_2= \frac{5p-3+2\nu_2(p-1)}{4(\nu_2-\nu_1)(p-1)}.
\end{eqnarray}
Notice that for $p\rightarrow\infty$ $\left.{\cal
P}_{\psi}(k,t)\right|_{IR}$ goes as $k^2$. An interesting result
for ${\cal P}_{\varphi}(k,t)$ is that it depends on $3p<
m^2=4(6\pi G \phi^2_0-1)<p(2p-1)$, and hence it is required that
$p>2$. On the other hand, for sufficiently large $t$ the first
term in (\ref{p1}) is dominant, so that
\begin{displaymath}
\left.{\cal P}_{\varphi}(k,t)\right|_{IR | t\rightarrow \infty}
\sim k^{3-2\nu_2},
\end{displaymath}
which approaches to a scale invariant spectrum for $p \rightarrow
\infty$.

\section{Final Remmarks}

In this work we have studied an example of the formalism developed
in \cite{kolb}, which is an extension of the well-known formalism
for gauge-invariant scalar metric fluctuations during inflation.
The formalism here examined is valid also for fluctuations with
large amplitude, but the equations are very difficult to be solved
due to the non-linearity of the Einstein and Lagrange equations.
In the proposal here studied vector and tensor perturbations of
the metric are neglected and the fluid is considered as
irrotacional. Of course, the analysis is only valid in a
cosmological context on super Hubble scales when the universe
expands adiabatically. We have confirmated that, for small
fluctuations the linear approximations give us the same dynamics
that for the standard method. In this work we have illustrated one
example where the universe grows with a scale factor $a(t) \sim
t^{p}$, (with $p \gg 1$). We found that, for very large $p$, at
the end of inflation the spectrum $\left.{\cal
P}_{\varphi}(k,t)\right|_{IR | t\rightarrow \infty}$ becomes scale
invariant on cosmological scales, but $\left.{\cal
P}_{\psi}(k,t)\right|_{IR | t\rightarrow \infty}$ goes as $k^2$.
However, at the beginning of inflation it is not true, because the
spectrum of the $\varphi$-fluctuations is altered by the modes of
metric fluctuations $\tilde\xi_k(t)$. Furthermore
$\left<E|\psi|E\right> \sim 1/a$, so that we conclude that at the
end of inflation (and later), next order effects
due to metric fluctuations on cosmological scales should be negligible. \\

\centerline{\bf{Acknowledgements}} \noindent The authors
acknowledge UNMdP and CONICET (Argentina) for
financial support.\\

\end{document}